\documentstyle[12pt]{article}

\catcode`\@=11
\long\def\@makefntext#1{
\protect\noindent \hbox to 3.2pt {\hskip-.9pt
$^{{\ninerm\@thefnmark}}$\hfil}#1\hfill}		

\def\@makefnmark{\hbox to 0pt{$^{\@thefnmark}$\hss}}  

\def\ps@myheadings{\let\@mkboth\@gobbletwo
\def\@oddhead{\hbox{}
\rightmark\hfil\ninerm\thepage}
\def\@oddfoot{}\def\@evenhead{\ninerm\thepage\hfil
\leftmark\hbox{}}\def\@evenfoot{}
\def\sectionmark##1{}\def\subsectionmark##1{}}

\renewcommand{\thefootnote}{\fnsymbol{footnote}}

\newcounter{sectionc}\newcounter{subsectionc}\newcounter{subsubsectionc}
\renewcommand{\section}[1] {\vspace*{0.6cm}\addtocounter{sectionc}{1}
\setcounter{subsectionc}{0}\setcounter{subsubsectionc}{0}\noindent
	{\normalsize\bf\thesectionc. #1}\par\vspace*{0.4cm}}
\renewcommand{\subsection}[1] {\vspace*{0.6cm}\addtocounter{subsectionc}{1}
	\setcounter{subsubsectionc}{0}\noindent
	{\normalsize\it\thesectionc.\thesubsectionc. #1}\par\vspace*{0.4cm}}
\renewcommand{\subsubsection}[1]
{\vspace*{0.6cm}\addtocounter{subsubsectionc}{1}
	\noindent {\normalsize\rm\thesectionc.\thesubsectionc.\thesubsubsectionc.
	#1}\par\vspace*{0.4cm}}

\newcounter{appendixc}
\newcounter{subappendixc}[appendixc]
\newcounter{subsubappendixc}[subappendixc]

\renewcommand{\appendix}[1] {\vspace*{0.6cm}
        \refstepcounter{appendixc}
        \setcounter{figure}{0}
        \setcounter{table}{0}
        \setcounter{equation}{0}
        \renewcommand{\thefigure}{\Alph{appendixc}.\arabic{figure}}
        \renewcommand{\thetable}{\Alph{appendixc}.\arabic{table}}
        \renewcommand{\theappendixc}{\Alph{appendixc}}
        \renewcommand{\theequation}{\Alph{appendixc}.\arabic{equation}}
        \noindent{\bf Appendix \theappendixc #1}\par\vspace*{0.4cm}}

\def\abstracts#1{{

\centering{\begin{minipage}{12.2truecm}\footnotesize\baselineskip=12pt\noindent
	\centerline{\footnotesize ABSTRACT}\vspace*{0.3cm}
	\parindent=0pt #1
	\end{minipage}}\par}}

\renewenvironment{thebibliography}[1]
	{\begin{list}{\arabic{enumi}.}
	{\usecounter{enumi}\setlength{\parsep}{0pt}
\setlength{\leftmargin 1.25cm}{\rightmargin 0pt}
	 \setlength{\itemsep}{0pt} \settowidth
	{\labelwidth}{#1.}\sloppy}}{\end{list}}

\newcounter{itemlistc}
\newcounter{romanlistc}
\newcounter{alphlistc}
\newcounter{arabiclistc}

\newcommand{\fcaption}[1]{
        \refstepcounter{figure}
        \setbox\@tempboxa = \hbox{\footnotesize Fig.~\thefigure. #1}
        \ifdim \wd\@tempboxa > 6in
           {\begin{center}
        \parbox{6in}{\footnotesize\baselineskip=12pt Fig.~\thefigure. #1}
            \end{center}}
        \else
             {\begin{center}
             {\footnotesize Fig.~\thefigure. #1}
              \end{center}}
        \fi}

\newcommand{\tcaption}[1]{
        \refstepcounter{table}
        \setbox\@tempboxa = \hbox{\footnotesize Table~\thetable. #1}
        \ifdim \wd\@tempboxa > 6in
           {\begin{center}
        \parbox{6in}{\footnotesize\baselineskip=12pt Table~\thetable. #1}
            \end{center}}
        \else
             {\begin{center}
             {\footnotesize Table~\thetable. #1}
              \end{center}}
        \fi}

\def\@citex[#1]#2{\if@filesw\immediate\write\@auxout
	{\string\citation{#2}}\fi
\def\@citea{}\@cite{\@for\@citeb:=#2\do
	{\@citea\def\@citea{,}\@ifundefined
	{b@\@citeb}{{\bf ?}\@warning
	{Citation `\@citeb' on page \thepage \space undefined}}
	{\csname b@\@citeb\endcsname}}}{#1}}

\newif\if@cghi
\def\cite{\@cghitrue\@ifnextchar [{\@tempswatrue
	\@citex}{\@tempswafalse\@citex[]}}
\def\citelow{\@cghifalse\@ifnextchar [{\@tempswatrue
	\@citex}{\@tempswafalse\@citex[]}}
\def\@cite#1#2{{$\null^{#1}$\if@tempswa\typeout
	{IJCGA warning: optional citation argument
	ignored: `#2'} \fi}}

 1
 1
 1

\font\ninerm=cmr9

\begin{document}

\newcommand{\st}{\scriptstyle}
\newcommand{\sst}{\scriptscriptstyle}
\newcommand{\mco}{\multicolumn}
\newcommand{\epp}{\epsilon^{\prime}}
\newcommand{\vep}{\varepsilon}
\newcommand{\ra}{\rightarrow}
\newcommand{\ppg}{\pi^+\pi^-\gamma}
\newcommand{\vp}{{\bf p}}
\newcommand{\ko}{K^0}
\newcommand{\kb}{\bar{K^0}}
\newcommand{\al}{\alpha}
\newcommand{\ab}{\bar{\alpha}}
\newcommand{\bean}{\begin{eqnarray*}}
\newcommand{\eean}{\end{eqnarray*}}
\newcommand{\hPi}{\mbox{$\widehat{\Pi}$}}
\newcommand{\hGa}{\mbox{$\widehat{\Gamma}$}}
\newcommand{\hT}{\mbox{$\widehat{T}$}}
\newcommand{\gi}{gauge-invariant}
\newcommand{\WI}{Ward identity}
\newcommand{\WIs}{Ward identities }
\newcommand{\eb}{\mbox{$\beta$}}
\newcommand{\eg}{\mbox{$\gamma$}}
\newcommand{\ez}{\mbox{$\zeta$}}
\newcommand{\el}{\mbox{$\lambda$}}
\newcommand{\es}{\mbox{$\sigma$}}
\newcommand{\er}{\mbox{$\rho$}}
\newcommand{\nn}{\nonumber}
\def\be{\begin{equation}}
\def\ee{\end{equation}}
\def\bea{\begin{eqnarray}}
\def\eea{\end{eqnarray}}
\def\CPbar{\hbox{{\rm CP}\hskip-1.80em{/}}}
\def \dsl {\partial \kern-.55em{/}}
\def \Dsl {D \kern-.65em{/}}
\def \qsl {q \kern-.45em{/}}
\def \slp {p \kern-.45em{/}}
\def \ksl {k \kern-.45em{/}}
\def \Gf {Green's function~}
\def \Gfs {Green's functions~}
\def \g5 {\gamma _5}

\centerline{\normalsize\bf GAUGE INVARIANCE AND}
\baselineskip=22pt
\centerline{\normalsize\bf ANOMALOUS GAUGE BOSON COUPLINGS}
\baselineskip=16pt
\vspace*{0.6cm}
\centerline{\footnotesize JOANNIS PAPAVASSILIOU}
\baselineskip=13pt
\centerline{\footnotesize\it Department of Physics, New York University,
4 Washington Place}
\baselineskip=12pt
\centerline{\footnotesize\it New York, NY 10003, USA}
\centerline{\footnotesize E-mail: papavass@mafalda.physics.nyu.edu}
\vspace*{0.3cm}
\centerline{\footnotesize and}
\vspace*{0.3cm}
\centerline{\footnotesize KOSTAS PHILIPPIDES}
\baselineskip=13pt
\centerline{\footnotesize\it Department of Physics, New York University,
4 Washington Place}
\baselineskip=12pt
\centerline{\footnotesize\it New York, NY 10003, USA}
\centerline{\footnotesize E-mail: kostas@mafalda.physics.nyu.edu}

\vspace*{0.9cm}
\abstracts{ Using the S--matrix pinch technique we obtain to one loop order,
gauge independent $\gamma W^-W^+$ and  $Z W^-W^+$ vertices in the context of
the standard model, with all incoming momenta off--shell. We show that the
vertices so constructed satisfy simple QED--like Ward identities.
These gauge invariant vertices give rise to
expressions for the
magnetic dipole and electric quadrupole
form factors of the W gauge boson,
which, unlike previous treatments,
satisfy the crucial properties of infrared finiteness and
perturbative unitarity.}

\normalsize\baselineskip=15pt
\setcounter{footnote}{0}
\renewcommand{\thefootnote}{\alph{footnote}}
\section{Introduction}
A new and largely unexplored frontier on which the ongoing search for new
physics will soon focus is the study of the structure of the three-boson
couplings.
A general parametrization of the trilinear gauge boson vertex for
on--shell $W$s and off--shell  $V=\gamma,~Z$ is
\bea
\Gamma_{\mu\alpha\beta}^{V}= & -ig_{V}~\Big[~
 ~f \left[~ 2g_{\alpha\beta}\Delta_{\mu}+ 4(g_{\alpha\mu}Q_{\beta}-
g_{\beta\mu}Q_{\alpha})~\right]
 +~ 2\Delta\kappa_{V}~(g_{\alpha\mu}Q_{\beta}-g_{\beta\mu}Q_{\alpha})
\nonumber \\
& ~~+~  4\frac{
{\Delta Q_{V}}}{{M_{W}^{2}}}
(\Delta_{\mu}Q_{\alpha}Q_{\beta}-
\frac{1}{2}Q^{2}g_{\alpha\beta}\Delta_{\mu})
{}~~\Big]~ +~ ... ~~~,
\eea
with $g_{\eg}= gs$, $g_{Z}= gc$, where $g$ is the  $SU(2)$
gauge coupling, $s\equiv sin\theta_{W}$ and $c\equiv cos\theta_{W}$,
 and the ellipses denote omission of C, P, or T violating terms.
 The four-momenta $Q$ and $\Delta$
 are related to the incoming momenta $q$, $p_{1}$ and $p_{2}$ of
the gauge bosons $V,~W^-$and $W^+$ respectively, by
$q=2Q$, $p_{1}=\Delta -Q$ and $p_{2}=-\Delta - Q$~
\cite{Bardeen}.
The form factors $\Delta\kappa_{V}$ and $\Delta Q_{V}$,
also defined as
$
\Delta\kappa_{V}= \kappa_{V} + \lambda_{V} - 1
$ ,
and
$
 \Delta Q_{V}= -2\lambda_{V}
$,
are compatible with C, P,
and T invariance, and
are  related to the magnetic dipole moment $\mu_{W}$ and the electric
quadrupole moment $Q_{W}$, by the following expressions:
\be
\mu_{W} = \frac{e}{2M_{W}}(2+ \Delta\kappa_{\gamma})
{}~~\mbox{,}~~
Q_{W}= -\frac{e}{M^{2}_{W}}(1+\Delta\kappa_{\gamma}+\Delta Q_{\gamma})~.
\ee
In the context of the standard model,
 their canonical, tree level values are
 $f=1$ and  $\Delta\kappa_{V}=\Delta Q_{V}=0$.
To determine the radiative corrections to these quantities
one must cast the resulting one--loop expressions in the following form:
\be
\Gamma_{\mu\alpha\beta}^{V}= -ig_{V}[
 a_{1}^{V}g_{\alpha\beta}\Delta_{\mu}+ a_{2}^{V}(g_{\alpha\mu}Q_{\beta}-
g_{\beta\mu}Q_{\alpha})
 + a_{3}^{V}\Delta_{\mu}Q_{\alpha}Q_{\beta}]~,
\label{1loopParametrization}
\ee
where $a_{1}^{V}$, $a_{2}^{V}$, and $a_{3}^{V}$ are
 complicated functions of the
momentum transfer $Q^2$, and the masses of the particles
appearing in the loops.
It then follows that  $\Delta\kappa_{V}$ and $\Delta Q_{V}$
are given by the
following expressions:
\be
\Delta\kappa_{V}=\frac{1}{2}(a_{2}^{V}-2a_{1}^{V}-Q^{2}a_{3}^{V})
{}~\mbox{,}~~
\Delta Q_{V}= \frac{M^{2}_{W}}{4}a_{3}^{V}~.
\label{1loopdeltakappaandQ}
\ee
Calculating the one-loop expressions for $\Delta\kappa_{V}$ and
 $\Delta Q_{V}$ is a non-trivial task, both from
the technical and the conceptual point of view.
If one
calculates
just the Feynman diagrams contributing to the $\gamma W^{+}W^{-}$
vertex and then extracts from them the contributions to
$\Delta\kappa_{\gamma}$ and $\Delta Q_{\gamma}$,
one arrives at
expressions that are
plagued with several pathologies, gauge-dependence being one of them.
Indeed, even if the two W are considered to be on shell, since the incoming
photon is not, there is no {\sl a priori}
 reason why a gauge-independent (g.i.) answer
should emerge. In the context of the renormalizable $R_{\xi}$ gauges
the final answer depends on the
choice of the gauge fixing parameter $\xi$, which enters into the one-loop
calculations through the gauge-boson propagators
( W,Z,$\gamma$, and unphysical Higgs particles).
In addition, as shown by an explicit calculation
performed
in the Feynman gauge ($\xi=1$), the answer for
$\Delta\kappa_{\gamma}$
is {\sl infrared divergent} and
violates perturbative unitarity,
e.g. it grows monotonically for $Q^2 \rightarrow \infty$
{}~\cite{Lahanas}.
All the above pathologies may be circumvented
if one adopts the pinch technique (PT)
\cite{Cornwall}.
The application of this method gives rise to
new expressions,
$\hat{\Delta}\kappa_{\gamma}$ and $\hat{\Delta} Q_{\gamma}$,
which are
gauge fixing parameter ($\xi$) independent,
ultraviolet {\sl and} infrared finite, and
well behaved for large momentum transfers $Q^{2}$.

\section{The pinch technique}
The S-matrix pinch technique is an algorithm that allows the construction
of modified g.i. $n$-point functions, through the
order by order rearrangement of
Feynman graphs, contributing to a certain physical
and therefore ostensibly g.i. amplitude,
(an S-matrix in our case). This rearrangement separates the S--matrix
into kinematically distinct pieces, akin to self--energies, vertices
and boxes,
 from which the new effective $n$-point functions can be
extracted. The resulting expressions are independent of the initial
gauge choice
and of the specific process one employs.
 For the case of vertices, the PT
 amounts into identifying and appending  to the usual vertex--graphs
those parts of the box graphs that exhibit a vertex--like structure.
These parts are called pinch parts;
they emerge every time a gauge--boson propagator, a three gauge
boson vertex or a scalar--scalar--gauge boson vertex contributes an
integration
momentum $k_{\mu}$ to the original graph's numerator.
Such momenta,
when contracted with a $\gamma$ matrix, trigger
 an elementary identity of the form
\be
k_{\mu}\gamma^{\mu} P_{L}\equiv \ksl P_{L} =
S_{i}^{-1}(p+k) P_{L} - P_{R}S_{j}^{-1}(p) + m_{i}P_{L} - m_{j}P_{R}~.
\label{BrokenPinch}
\ee
The first term on the r.h.s. of Eq.(\ref{BrokenPinch}) removes the
internal fermion propagator and generates a pinch term, while
the second vanishes on shell. Collecting all vertex-like
pinch parts and appending them to the usual vertex graphs
gives rise to a g.i. sub--amplitude, with the same kinematical properties
as a vertex.

\section{Gauge--invariant gauge boson vertices
 and their Ward identities}

We consider
the S-matrix element for the process
\begin{equation}
e^-  + \nu  +  e^- \to
e^- + e^-  +  {\overline \nu}     ~~.
\label{process}
\end{equation}
and isolate the part $T(q,p_1,p_2)$ of the S--matrix which depends
only on the momentum transfers $q$, $p_1$, and $p_2$.
Since the final result (with pinch contributions included)
is g.i., we choose to work in the Feynman gauge
($\xi_{i}=1$); this particular gauge simplifies the calculations
because it removes all longitudinal parts from the
tree-level gauge boson propagators. So, pinch contributions can
only originate from appropriate momenta furnished by the
tree--level gauge boson vertices.
Applying the pinch technique algorithm we isolate all
vertex--like parts contained
in the box diagrams and allot them to the usual vertex
graphs.
The final expressions for
one loop g.i. trilinear gauge boson vertices are :
\bea
\frac{1}{g^3s}
\hGa^{\gamma W^{-}W^{+}} _{\mu \alpha \beta} & =&
\Gamma^{\gamma W^-W^+}_{\mu \alpha \beta} |_{\xi _i =1}
 + q^{2}~B_{\mu\alpha \beta}
 +U_W^{-1}(p_1)^{\rho}_{\alpha} B^+_{\mu \rho \beta}
  +U_W^{-1}(p_2)^{\rho}_{\beta} B^-_{\mu \alpha \rho}  \nonumber \\
&& - 2 \Gamma_{\mu\alpha\beta} \left[~
I_{WW}(q)+s^2I_{W\gamma}(p_1)+c^2I_{WZ}(p_1)+s^2I_{W\gamma}(p_2)
+c^2I_{WZ}(p_2) ~\right]
\nonumber \\
&& + p_{2 \beta} g_{\mu \alpha} ~ {\cal M}^-
  +p_{1 \alpha} g_{\mu \beta}~  {\cal M}^+ ~~,
\label{gigWW}
\eea
\bea
\frac{1}{g^3c}
\hGa^{ZW^{-}W^{+}} _{\mu \alpha \beta} & =&
\Gamma^{Z W^-W^+}_{\mu \alpha \beta} |_{\xi _i =1}
+ U_Z^{-1}(q)^{\rho}_{\mu} B_{\rho \alpha \beta}
 +U_W^{-1}(p_1)^{\rho}_{\alpha} B^+_{\mu \rho \beta}
  +U_W^{-1}(p_2)^{\rho}_{\beta} B^-_{\mu \alpha \rho}  \nonumber \\
 && -  2 \Gamma_{\mu\alpha\beta} \left[~
I_{WW}(q)+s^2I_{W\gamma}(p_1)+c^2I_{WZ}(p_1)+s^2I_{W\gamma}(p_2)
+c^2I_{WZ}(p_2) ~\right] \nonumber \\
&& + q_{\mu} g_{\alpha \beta}~M_Z^2~ {\cal M}
+ p_{2 \beta} g_{\mu \alpha} ~M_W^2~ {\cal M}^-
  + p_{1 \alpha} g_{\mu \beta}~M_W^2~  {\cal M}^+
\label{giZWW} ~~.
\eea
The quantities $I_{ij}$ and ${\cal M}^{\pm}$
are integrals with two and three scalar propagators, respectively;
their explicit expressions
have been reported elsewhere
\cite{PaPhi}.

The g.i. vertices satisfy the following simple Ward
identities (WI), relating them to the
g.i. $W$ self energy and $\chi WW$ vertex constructed also via the PT :
\be
q^\mu \hGa ^{ZW^-W^+} _{\mu \alpha \beta}
+ iM_Z \hGa ^{\chi W^-W^+} _{\alpha \beta}
 =  gc ~\left[ \hPi ^W _{\alpha \beta} (1) - \hPi ^W _{\alpha \beta} (2)
\right]~~,
\label{qZWW}
\ee
\be
q^\mu \hGa ^{\gamma W^-W^+} _{\mu \alpha \beta}
 =  gs ~\left[ \hPi ^W _{\alpha \beta} (1) - \hPi ^W _{\alpha \beta} (2)
\right] ~~.
\label{qZWW}
\ee
These WI
 are the one--loop generalizations of the respective {\it tree level} WI;
their validity is
crucial for the gauge independence of the S--matrix.
It is important to emphasize that they make no
reference to ghost terms, unlike the corresponding Slavnov-Taylor
identities satisfied by the conventional, gauge--dependent vertices.

For the case of {\it on--shell}
{}~$W$s one sets $p_1^2=p_2^2=M^2_W$ and neglects all
terms proportional to $p_{1\alpha}$ and $p_{2 \beta}$,
as well as the left over pinch terms of the $W$ legs. Then the
$\gamma WW$ vertex reduces to the form
\be
\frac{1}{g^3s}
\hGa^{\gamma W^{-}W^{+}} _{\mu \alpha \beta}  =
\Gamma^{\gamma W^-W^+}_{\mu \alpha \beta} |_{\xi _i =1}
 + q^{2}~B_{\mu\alpha \beta}(q,p_1,p_2)
 - 2 \Gamma_{\mu\alpha\beta}
I_{WW}(q)~~.
\label{gigWW}
\ee
This is of course the same answer one obtains by applying the PT
{\it directly} to the S--matrix of $e^{+}e^{-} \rightarrow W^{+}W^{-}$.
Thus for the form factors ${\Delta\kappa}_{\gamma},~{\Delta Q}_{\gamma}$
the only function we need is $B_{\mu\alpha \beta}$, given below
\be
g^2 B_{\mu\alpha \beta} = \sum_{V=\gamma Z} g_{V}^2
\int\frac{d^4k}{i(2\pi)^4}
\frac{g_{\alpha \beta} ~(k-\frac{3}{2}(p_1-p_2))_{\mu}
-g_{\alpha \mu}~(3k+2q)_{\beta}
-g_{\beta \mu}~(3k-2q)_{\alpha}}
{\left[(k+p_1)^2-M_W^2\right]\left[(k-p_2)^2-M_W^2\right]
\left[k^2-M_V^2\right]}
{}~.
\ee

\section{Magnetic dipole and electric quadrupole form factors for the $W$}

Having constructed the g.i. $\gamma WW$ vertex we proceed to extract
its contributions to the magnetic dipole and electric quadrupole form
factors of the $W$.
We use carets to denote the g.i. one--loop contributions.
Clearly,
\be
\hat{\Delta}\kappa_{\gamma} = \Delta\kappa_{\gamma}^{(\xi=1)}
+ \Delta\kappa_{\gamma}^{P}~,
\ee
and
\be
\hat{\Delta}Q_{\gamma} = \Delta Q_{\gamma}^{(\xi=1)}+  \Delta Q_{\gamma}^{P} ~.
\ee
where $\Delta Q_{\gamma}^{(\xi=1)}$ and $\Delta Q_{\gamma}^{(\xi=1)}$
are the contributions of the usual vertex diagrams in
the Feynman gauge \cite{Lahanas}, whereas
$\Delta Q_{\gamma}^P$ and $\Delta Q_{\gamma}^P$the analogous
contributions from the pinch parts.
By performing the momentum integration in $B_{\mu\alpha\beta}$, we find for
$p_1^2=p_2^2=M_W^2$
\be
B_{\mu\alpha\beta}= -\frac{Q^2}{8\pi^2 M^2_W}\sum_{V=\gamma, Z} g^2_{V}
\int_{0}^{1}da \int_{0}^{1}(2tdt) \frac{F_{\mu\alpha\beta}}{L_{V}^{2}}~,
\ee
where
\be
F_{\mu\alpha\beta}
= 2(\frac{3}{2} + at)g_{\alpha\beta}{\Delta}_{\mu}+
2(3at+2)[g_{\alpha\mu}Q_{\beta}-g_{\beta\mu}Q_{\alpha}]~,
\label{F}
\ee
and
\be
L^{2}_{V} = t^{2}-t^{2}a(1-a)(\frac{4Q^{2}}{M^{2}_{W}}) +
 (1-t)\frac{M^{2}_{V}}{M^{2}_{W}}~,
\ee
from which follows that
\be
{\Delta\kappa}_{\gamma}^{P}= -\frac{1}{2} \frac{Q^{2}}{M^{2}_{W}}
\sum_{V} \frac{\alpha_{V}}{\pi} \int_{0}^{1}da \int_{0}^{1}(2tdt)
\frac{(at-1)}{L^{2}_{V}}~,
\ee
and
\be
{\Delta Q}_{\gamma}^{P} = 0 ~~~.
\ee
We observe that $\Delta\kappa_{\gamma}^{P}$ contains an
infrared divergent term, stemming from the double
integral shown above, when $V=\gamma$. This term cancels exactly
against a similar infrared divergent piece
contained in $\Delta\kappa_{\gamma}^{(\xi = 1)}$,
{}~\cite{Lahanas} thus rendering
$\hat{\Delta}\kappa_{\gamma}$ infrared finite.
After the infrared pieces have been cancelled, one notices that
the remaining contribution of $\Delta\kappa_{\gamma}^{P}$
decreases monotonically
as $Q^2 \rightarrow \pm \infty$; due to the difference
in relative signs this contribution cancels
asymptotically against the monotonically increasing
contribution from
$\Delta\kappa_{\gamma}^{(\xi = 1)}$.
Thus by including the pinch part the
unitarity of $\hat{\Delta}\kappa_{\gamma}$ is restored and
$\hat{\Delta}\kappa_{\gamma} \rightarrow 0$ for large values of $Q^2$.
It would be interesting to determine how these quantities could be
directly extracted from future $e^{+}e^{-}$ experiments.

\section{Conclusions}
We showed how to use the PT in order to construct
g.i.
gauge--boson vertices,
which satisfy naive QED--like Ward identities.
These vertices give rise to
magnetic dipole and electric quadrupole form factors for the $W$,
which can, at least in principle, be promoted to physical
observables.

\section {Acknowledgment}
The authors are happy to acknowledge useful discussions with
J.~M.~Cornwall, A.~Lahanas, C.~Papadopoulos,
A.~Sirlin, and D.~Zeppenfeld.
This work was supported in part by the National Science Foundation under
Grant No.PHY-9017585.


\end{document}